\documentstyle[aps,prb,epsf,twocolumn]{revtex}

\draft

\def\bra{\langle}
\def\ket{\rangle}
\def\vA{{\bf A}}
\def\vp{{\bf p}}
\def\vr{{\bf r}}
\def\vrp{\vr^\prime}
\def\vv{{\bf v}}
\def\nvr{{\bf r}_1,\cdots{\bf r}_N}
\def\dnvr{\left(\prod_{i=1}^N\int d^2{\bf r}_i\right)}

\def\DE{\Delta\varepsilon}
\def\dfrac{\displaystyle\frac}

\def\op{\Theta}
\def\opi{\Theta_I}
\def\rp{r^\prime}
\def\phip{\phi^\prime}
\def\thep{\theta^\prime}
\def\Deth{\Delta\theta}

\def\gsw{\Psi_{\rm gs}}
\def\cvw{\Psi_{\vr_0}}
\def\cvwp{\Psi_{\vrp_0}}

\def\weighting{\int_0^R\!dr_0r_0f_{n,l}(r_0)\int_0^R\!d\rp_0\rp_0f_{n,l}(\rp_0)}

\begin{document}

\wideabs{
\title{Cyclotron motion of a quantized vortex in a superfluid}
\author{Jian-Ming Tang}
\address{Department of Physics, University of Washington, Box 351560, Seattle WA 98195-1560, USA}

\maketitle

\begin{abstract}

In two dimensions a microscopic theory providing a basis for the naive
analogy between a quantized vortex in a superfluid and an electron in
a uniform magnetic field is presented. Following the variational
approach developed by Peierls, Yoccoz, and Thouless, the cyclotron
motion of a vortex is described by the many-body wave function, which
is a linear combination of Feynman wave functions centered at
different positions. An integral equation for the weighting functions
of the superposition is derived by minimizing the energy
functional. The matrix elements of the kernel are the overlaps between
any two displaced Feynman wave functions. A numerical study is
conducted for a bosonic superfluid based on a Hartree ground state.  A
one-to-one correspondence between the rotational states of a vortex in
a cylinder and the cyclotron states of an electron in the central
gauge is found. Like the Landau levels of an electron, the energy
levels of a vortex are highly degenerate. However, the gap between two
adjacent energy levels does not only depend on the quantized
circulation, but also increases with energy, and scales with the size
of the vortex. The fluid density is finite at the vortex axis and the
vorticity is distributed in the core region. The effective mass of a
quantized vortex defined by the inverse of the energy-level spacing is
shown to be logarithmically divergent with the size of the vortex.

\end{abstract}


}

\section{Introduction}

Quantized vortices have been known to play a significant part in the
behaviors of a superfluid for several decades since
Onsager~\cite{Onsager49} and Feynman.~\cite{Feynman55} Understanding
the vortex dynamics from the quantum theoretical point of view still
remains as a challenging problem after these years. Recently, an
analogy with the cyclotron motion of an electron has helped us to make
a few progresses on this subject.

The phenomenological theory,~\cite{Hatsuda94} which models a quantized
vortex in a two-dimensional superfluid as a charged particle in a
gauge field, has been very intriguing for understanding the dynamics
of the vortex. This analogy is motivated by the fact that, in
classical physics, the motion of a vortex in an ideal fluid is similar
to the motion of an electron in an uniform magnetic field. In both
cases the broken time reversal symmetry leads to a transverse force
proportional to the velocity. The Magnus force due to the Bernoulli
pressure difference is equal to the mass density of the fluid times
the vector product between the circulation vector along the vortex
axis and the velocity of the vortex relative to the background
fluid. The well-known Lorentz force is equal to the charge of the
electron times the vector product between the magnetic field and the
velocity of the electron relative to the field. Following the simple
relationship between the energy and the momentum of an ideal fluid, a
vortex is like a particle with an effective mass.~\cite{Lamb32} The
contributions to the effective mass come both from the particles
trapped inside the vortex core and from the fluid disturbed by the
motion of the vortex. As a result of the velocity-dependent transverse
force, the trajectory of a classical vortex consists of a circular
orbit around a guiding center with an arbitrary radius but a fixed
angular frequency.  In the superfluid $^4$He, the dynamics of a
quantized vortex is also governed by the Magnus force and the
effective mass because the compressibility and the viscosity of the
fluid are comparably small.~\cite{Donnelly91}

Having the strong analogy in classical physics between an electron and
a vortex in mind, one expects to find some common features in their
fully quantized theories. However, the actual theoretical developments
of these two systems are quite different from one another. On one
hand, enormous progress has been made on the quantum theory of
electrons in the presence of a high magnetic field since the discovery
of the quantum Hall effect.~\cite{VonKlitzing80} Even though the real
system consists of strongly correlated electrons, the basic physics of
the integer quantum Hall effect can be understood from the simple
theory of a non-interacting electron gas on a two-dimensional
plane. In order to compare with the vortex motion later, I will
discuss the solution in the central gauge so that the angular momentum
in the perpendicular direction to the plane is conserved. In the
central gauge, electrons move as if in a rotating frame with an
effective simple-harmonic potential. The energy eigenstates shift up
and down from the simple harmonic spectrum according to their
individual angular momentum, and form highly degenerate Landau
levels. The detailed solution to this idealized problem without any
disorder is given in the appendix of this paper. On the other hand,
there is no quantum theory starting from first principles which
supports the existence of the cyclotron motion and degenerate levels
of vortex states. Most theories of vortex dynamics are semi-classical
in the sense that the motion of a vortex is described explicitly by
the coordinates of the vortex center moving on a classical
trajectory. The governing equations for the vortex center and the
quantization conditions are not known due to a fundamental difference
between an electron and a vortex: an electron is a real particle
obeying Schr\"odinger equation, but a vortex is a collective object in
a many-body system. Experimental evidences for these collective modes
in the superfluid $^4$He have been shown in a more complicated
situation, where the cyclotron motion is coupled to a traveling wave
along the vortex axis.~\cite{Ashton79} In high-$T_c$ superconductors,
the vortex cyclotron resonance has also been studied in several recent
experiments. \cite{Drew95,Ichiguchi98}

It is the purpose of the present paper to provide a microscopic theory
for this phenomenological analogy, and show that the rotational states
of a quantized vortex also form highly degenerate energy levels
similar to the Landau levels in the integer quantum Hall effect. In
order to describe these rotational states, which are collective modes
in a many-body system, I construct variational wave functions using
the projection method adapted from the nuclear theory. It was
suggested by Hill and Wheeler~\cite{Hill53} that the wave functions of
rotational states in a nucleus could be constructed by integrating out
the surface variables, which specify the position and the orientation
of the nucleus. The wave functions of a quasi-harmonic oscillator were
proposed for the weighting functions of the integration. Peierls,
Yoccoz and Thouless~\cite{Peierls57,Peierls62} later pointed out that
the weighting functions could be determined from the variational
principle. I apply their scheme on a vortex in a bosonic superfluid,
and calculate the energy spectrum and the weighting functions in the
dilute-gas limit. In certain sense these weighting functions can be
understood as the ``single-particle wave functions'' of the vortex. At
the end of this paper, the weighting functions are shown numerically
to be identical to the eigenfunctions of a rotating two-dimensional
harmonic oscillator, which is consistent with the analogy of cyclotron
motion. Degenerate energy levels like Landau levels are also found in
this theory. However, the energy gap between two neighboring levels is
not a constant, and scales logarithmically with the size of the
vortex.

In Sec.~\ref{sc} a brief review is given on the semi-classical wave
functions of a quantized vortex in a superfluid. These many-body wave
functions are constructed with the help of classical hydrodynamics,
and will be used as the basis functions for constructing the new trial
wave functions.

In Sec.~\ref{proj} the projection method is discussed in terms of the
quantum fluctuation of the vortex position. I propose the new set of
trial wave functions based on the linear combination of the
semi-classical wave functions. An integral equation for the weighting
function of the superposition is obtained by using the variational
principle.

In Sec.~\ref{har} a system of bosons with short-ranged repulsive
interactions is considered. In the dilute-gas limit, the ground state
of the system can be described by the Hartree approximation. The
kernel of the integral equation is reduced to a sum of products of
one-particle integrals.

In Sec.~\ref{sol} the exact solutions of the dilute system is obtained
by solving the integral equation numerically. Energy levels formed by
lots of degenerate states are found. A partial explanation for the
degeneracy is given. To visualize the structure of a vortex, I
calculate the radial distributions of the number density and the
current density. Since a simplified ground state is used, the curves
only show qualitative features comparing to the results of quantum
Monte Carlo calculations,~\cite{Ortiz95,Giorgini96,Vitiello96,Sadd97}
which, however, are only available for a static vortex.

In Sec.~\ref{mass} an excellent agreement between the vortex motion
and the cyclotron motion is found by comparing the weighting functions
and the electron wave functions. In analogy to the cyclotron motion,
the effective mass of a vortex is link to the inverse of the energy
gap, which scales logarithmically with the size of the vortex.

\section{Semi-Classical Wave functions}
\label{sc}

At the zero temperature, the wave function for an uniform flow in a
superfluid can be obtained by performing a Galilean transformation on
the ground state,
\begin{equation}
\Psi_{\vv_{\rm s}}=\exp\left(i\frac{m}{\hbar}\vv_{\rm s}\cdot\sum_{j=1}^N\vr_j\right)\gsw(\nvr) \;, \label{eq:uf}
\end{equation}
where $m$ is the mass of the particle, $\vv_{\rm s}$ is the flow
velocity, $\vr_j$ is the position vector of each particle, $N$ is the
total number of particles, and $\gsw$ is the ground state wave
function. The ground state wave function is chosen to be real and
positive apart from an arbitrary phase factor.~\cite{Feynman53} The
wave function in Eq.~(\ref{eq:uf}) can be generalized to represent a
flow slowly varying in space,
\begin{equation}
\Psi_{\vv_{\rm s}(\vr)}=\left[\prod_{j=1}^N e^{i\phi(\vr_j)}\right]\gsw(\nvr) \;, \label{eq:scwf}
\end{equation}
where the velocity field is described by the gradient of the phase of
the wave function,
\begin{equation}
\vv_s(\vr)=\frac{\hbar}{m}\nabla\phi(\vr) \;.
\end{equation}
Consequently, the motion of the superfluid has to be a potential flow
in a simply connected region.

In a rotating superfluid, vortices with quantized circulations are
formed. A naive picture of a vortex is a core with non-vanishing
vorticity immersed in an irrotational fluid.~\cite{Lamb32} For a
single static vortex fixed at a point $\vr_0$ in an infinite
homogeneous system, the velocity field outside the core is inversely
proportional to the distance,
\begin{equation}
\vv_s(\vr)=\frac{\hbar}{m|\vr-\vr_0|}\hat\theta(\vr-\vr_0) \;,
\end{equation}
where $\hat\theta(\vr-\vr_0)$ is the unit vector in the azimuthal
direction relative to the point $\vr_0$. In this region a wave
function of the form in Eq.~(\ref{eq:scwf}) can be constructed with
the phase function $\phi(\vr)$ equal to the azimuthal angle
$\theta(\vr-\vr_0)$. Inside the core, this construction scheme fails
completely because the velocity field is no longer curl-free, and
cannot be incorporated into a phase function with a single
variable. Moreover, the actual structure of the core is in general not
known. In the superfluid $^4$He, the size of the core is about the
size of a helium atom or less. Needless to say, it is very difficult
to probe the core structure at this scale
experimentally. Theoretically, the simplest remedy for constructing a
trial wave function is to assume that the phase function possess the
same form everywhere, but the amplitude of the wave function is
reduced around the core,
\begin{equation}
\cvw=\left[\prod_{j=1}^Ng(|\vr_j-\vr_0|)e^{i\theta(\vr_j-\vr_0)}\right]\gsw(\nvr) \;, \label{eq:Fw}
\end{equation}
where $g(r)$ is a cut-off function, which goes to zero at the origin,
and goes to one asymptotically. Eq.~(\ref{eq:Fw}) was first discussed
by Feynman,~\cite{Feynman55} and was extensively used in literatures
with the cut-off function determined variationally.\cite{Chester68}
This wave function is semi-classical in the sense that the vortex is
exactly at the point $\vr_0$, like a $\delta$-function in quantum
mechanics. Nevertheless, it is a fairly good description for a vortex
in the superfluid $^4$He because the uncertainty in the vortex
position is on the order of the inter-atomic spacing.

In the situation where the vortex is moving with a velocity $\vv_0$,
there is an additional dipolar flow generated by the motion of the
core. The exact flow pattern depends on the detail of the core. For
example, the velocity potential for a vortex with a hard cylindrical
core is
\begin{equation}
\phi(\vr)=\theta(\vr-\vr_0)-\frac{ma^2\vv_0\cdot(\vr-\vr_0)}{\hbar|\vr-\vr_0|^2}\;, \label{eq:bf}
\end{equation}
where $a$ is the radius of the hard core.~\cite{Lamb32} The additional
dipolar flow given by the second term in the above equation is called
backflow. The backflow is quite important for considering the
dynamical properties of the vortex. In classical hydrodynamics the
backflow is responsible for the effective mass of a vortex. One can
compute the total kinetic energy of the fluid from the velocity
field. The first-order term in $\vv_0$ vanishes, and the
coefficient of the second-order term gives the effective mass, which
in two dimensions is equal to the amount of the fluid displaced by the
hard core,
\begin{equation}
M_{\rm eff}=\rho\pi a^2 \;,
\end{equation}
where $\rho$ is the mass density of the fluid. In the superfluid
$^4$He, the backflow needs to be included in the trial wave functions
in order to obtain a quantitative agreement between calculations and
experimental data on the energy spectrum of quasi-particle
excitations, especially on the part of rotons.~\cite{Feynman56} The
rotons are usually visualized as vortex rings.

In the rest of this paper I have completely neglected all the backflow
corrections in the trial wave functions for the technical
difficulty. As a result, only qualitative features of the energy
spectrum and of the effective mass can be obtained. A few
papers~\cite{Ortiz95,Giorgini96} have included the backflow in the
trial wave function, but only the static vortex is discussed there. In
the static case, the correction is small because the vortex is not
really moving besides the zero-point motion.

\section{Projection Method}
\label{proj}

In the Feynman wave function a vortex is described both by the
coordinates of particles and by the coordinates of the vortex core. An
intuitive way to consider the vortex motion is to make the coordinates
of the core time-dependent like real dynamical variables. However,
like the coordinates for the center of mass, the coordinates of the
core are collective variables which do not show up explicitly in the
underlying microscopic Hamiltonian. A set of dynamical variables
including both the particle coordinates and the core coordinates is a
redundant set. In the absence of external pinning centers, the motion
of the vortex must have been accounted in a proper quantum mechanical
description of the $N$-particle system. How does the motion of the
vortex appear as part of the motion of a superfluid rather than as
something imposed from outside? Conversely, how can the motion of the
vortex be described by degrees of freedom which are not real dynamical
variables? Similar questions were asked long ago in a different
context regarding the rotational states of a large deformed nucleus. A
large nucleus is a many-body system which under certain conditions can
move or rotate as a whole. In the framework of the Hartree-Fock
theory, the wave function of a nucleus is written in terms of a Slater
determinant of one-particle wave functions in a self-consistent
potential well. If the translation or the rotation of the nucleus is
considered through the position or the orientation of the potential
well, then one also has a redundant set of dynamical variables.

The answer to the above questions was first given by Hill and Wheeler
in terms of the quantum fluctuation of the potential well. In the
context of vortex dynamics, the corresponding answer is the
fluctuation in the vortex position. In a collective motion, the
particle variables and the collective variables are strongly
correlated. Since there are quantum fluctuations in the motion of the
particles in a superfluid, there must be fluctuations in the position
of the vortex, too. Mathematically, we integrate out the fluctuating
variables other than the particle coordinates, and obtain a linear
combination of Feynman wave functions with given core coordinates,
\begin{equation}
\Psi=\int d^2\vr_0G(\vr_0)\cvw \;, \label{eq:trial}
\end{equation}
where the weighting function $G(\vr_0)$ represents the quantum
fluctuation. This wave function depends only on the coordinates of
particles, and in the mean time provides a description of the vortex
motion through the weighting function. $\{\vr_0\}$ is called generator
coordinates, and is a good set of independent coordinates in the
Hilbert space because any two Feynman wave functions centered at
different positions are nearly orthogonal to each other. In this case,
a naive guess for the weighting function is the wave function of a
rotating two-dimensional harmonic oscillator. I will show later that
this guess is indeed the correct one.

An alternative point of view to Eq.~(\ref{eq:trial}) was later
provided by Peierls, Yoccoz and Thouless, who also developed a method
to determine the weighting function systematically. A vortex located
at a certain point in space breaks the translational symmetry
explicitly. Therefore we have a family of semi-classical wave
functions parametrized by their center coordinates, and all of them
give the same expectation value in energy. Generally speaking, the
variational energy can be further reduced by using a superposition of
these degenerate functions as the new trial wave function, and the
weighting function can be determined by minimizing the energy
integral,
\begin{equation}
\bra E\ket=\frac{\int\int d^2\vrp_0d^2\vr_0 G^*(\vrp_0)G(\vr_0)\bra\cvwp|H|\cvw\ket}{\int\int d^2\vrp_0d^2\vr_0 G^*(\vrp_0)G(\vr_0)\bra\cvwp|\cvw\ket} \;,
\end{equation}
where $H$ is the Hamiltonian. The trial wave function can also regain
the proper symmetry of the Hamiltonian with a suitable choice of the
weighting function. This approach is then called the projection method
because it projects out a state with the desired symmetry.

I start with a system of $N$ bosons in a disk geometry with radius
$R$. The generic Hamiltonian is
\begin{equation}
H=-\frac{\hbar^2}{2m}\sum_{i=1}^N\left[\nabla_i^2+V_{\rm b}(r_i)\right]+\frac{1}{2}\sum_{i,j=1}^NV(|\vr_i-\vr_j|) \;,
\end{equation}
where the particles interact with one another through the potential
$V(r)$, and the boundary of the system is represented by the confining
potential $V_{\rm b}(r)$, which is a step function with an infinite
height. The explicit form of the inter-particle potential $V(r)$ is
not directly relevant to our discussion here, which will become clear
in a moment.

The first step is to find the velocity potential of a single vortex
located at $\vr_0=(x_0,y_0)$. This problem is solved by using an image
vortex with the opposite circulation located at a distance $R^2/r_0$
from the origin,~\cite{Saffman95}
\begin{equation}
\phi(\vr_j;\vr_0)=\tan^{-1}\frac{y_j-y_0}{x_j-x_0}-\tan^{-1} \frac{y_j-R^2y_0/r_0^2}{x_j-R^2x_0/r_0^2} \;. \label{eq:phase}
\end{equation}
One can easily verify that Eq.~(\ref{eq:phase}) is the solution by
checking that the radial current vanishes on the boundary. Because of
the cylindrical symmetry of the system, the weighting function in
Eq.~(\ref{eq:trial}) must be of the form,
\begin{equation}
G(\vr_0)=f_{n,l}(r_0)e^{il\theta_0} \;,
\end{equation}
where $n$ is the principal quantum number, and $l$ is the angular
momentum. Following the convention of Feynman,~\cite{Feynman54} I can
rewrite the trial wave function in the following form,
\begin{equation}
\Psi_{n,l}=F_{n,l}(\nvr)\gsw(\nvr) \;, \label{eq:simpleform}
\end{equation}
where all the phase factors are combined together,
\begin{eqnarray}
F_{n,l} & = & \int d^2\vr_0 f_{n,l}(r_0)e^{i\Phi_l(\nvr;\vr_0)} \;,\\
\Phi_l & = & l\theta_0+\sum_{j=1}^N\phi(\vr_j;\vr_0) \;.
\end{eqnarray}
For simplicity, I have set the cut-off function $g(r)$ in
Eq.~(\ref{eq:Fw}) to be unity. In this case, even though the Feynman
wave function $\cvw$ becomes singular, the trial function in
Eq.~(\ref{eq:trial}) remains to be regular. The cut-off function
cannot be set to unity if one wants to include the backflow in the
trial wave function. The expectation value of the Hamiltonian relative
to the ground state energy $E_{\rm gs}$ is
\begin{eqnarray}
\varepsilon_{n,l} & = & \bra\Psi_{n,l}|H|\Psi_{n,l}\ket-E_{\rm gs} \nonumber\\
& = & \dfrac{\dfrac{\hbar^2}{2m}\dnvr\sum_{j=1}^N|\nabla_j F_{n,l}|^2\gsw^2}{\displaystyle\dnvr |F_{n,l}|^2\gsw^2} \;.
\end{eqnarray}
It is a generic feature for trial wave functions of the form in
Eq.~(\ref{eq:simpleform}) that the inter-particle potential does not
show up explicitly in the variational energy $\varepsilon_{n,l}$. The
influences from the inter-particle interactions only come in
indirectly through the ground-state wave function. By varying the
weighting function $f_{n,l}(\rp_0)$ to minimize $\varepsilon_{n,l}$, I
obtain an integral equation for $f_{n,l}(r_0)$,
\begin{equation}
\int_0^Rdr_0\left[K_l(\rp_0,r_0)-\varepsilon_{n,l}J_l(\rp_0,r_0)\right]f_{n,l}(r_0)=0 \;, \label{eq:inteq}
\end{equation}
where the kernel of this integral equation contains two parts: $J_l$
is the angular average of the overlap between any two Feynman wave
functions located at the radius $r_0$ and $\rp_0$, and $K_l$ is the
similar overlap weighted by the kinetic energy,
\begin{eqnarray}
J_l(\rp_0,r_0) & = & r_0\!\int_{-\pi}^\pi\!\! d\Deth_0\dnvr e^{i(\Phi_l-\Phi_l^\prime)}\gsw^2 \;,\!\!\!\label{eq:ker1}\\
K_l(\rp_0,r_0) & = & \frac{\hbar^2}{2m}r_0\!\int_{-\pi}^\pi\!\! d\Deth_0\dnvr e^{i(\Phi_l-\Phi_l^\prime)} \nonumber\\
&& \times\left[\sum_{j=1}^N\nabla_j\phip(\vr_j)\cdot\nabla_j\phi(\vr_j)\right]\gsw^2 \;,\label{eq:ker2}
\end{eqnarray}
where the primed functions $\Phi^\prime_l$ and $\phip_l$ come from the
complex conjugate of the trial wave function, and are referred to the
corresponding unprimed functions with $\vr_0$ substituted by
$\vrp_0$. It is clear that the integrand only depends on the relative
angle, $\Deth_0=\thep_0-\theta_0$. All particle coordinates in the
integrand are equivalent as a result of the Bose symmetry.

\section{Hartree Approximation}
\label{har}

In order to simplify the kernel further and obtain an analytic form,
one needs an explicit form of the ground-state wave function. In the
Hartree approximation for bosons, the ground-state wave function is
broken down to a product of normalized one-particle wave functions,
\begin{equation}
\gsw\approx\prod_{i=1}^Nu(\vr_i) \;,
\end{equation}
where $u(\vr)$ satisfies the non-linear Schr\"odinger equation with a
self-consistent potential,
\begin{equation}
V_{\rm sc}(\vr)=(N-1)\int d^2\vrp V(|\vr-\vrp|)u(\vrp)^2+V_{\rm b}(\vr) \;.
\end{equation}
For a dilute system of bosons with short-ranged repulsive
interactions, this one-particle wave function is approximately
constant except for a transition layer near the boundary. Assuming
that the effective repulsive interactions is infinitely strong, I can
neglect the transition layer, and set my ground state to a
constant. Since the higher-order correlations among particles are
completely neglected in the Hartree approximation, the $N$-particle
multiple integral in Eqs.~(\ref{eq:ker1}) and (\ref{eq:ker2}) is
reduced to a simple product of one-particle integrals,
\begin{eqnarray} 
J_l(\rp_0,r_0) & = & r_0\!\int_{-\pi}^\pi\! d\Deth_0e^{-il\Deth_0}\left[\frac{1}{A}\!\int d^2\vr e^{i(\phi-\phip)}\right]^N \!\!,\!\!\! \label{eq:kers1}\\
K_l(\rp_0,r_0) & = & r_0\!\int_{-\pi}^\pi\! d\Deth_0e^{-il\Deth_0}\left[\frac{1}{A}\!\int d^2\vr e^{i(\phi-\phip)}\right]^{N-1} \nonumber\\
&& \times\frac{\hbar^2}{2m}\frac{N}{A}\int d^2\vr(\nabla\phip\cdot\nabla\phi)e^{i(\phi-\phip)} \;,\label{eq:kers2}
\end{eqnarray}
where the normalized ground state is $A^{-N/2}$, and $A$ is the area
of the system.

From now on, I will set the unit length to be the inter-particle
spacing, $\sigma=R/\sqrt{N}$, which is the only length scale in this
model. Other length scales can only come in implicitly through the
spatial variations of the ground-state wave function, and are
discarded in my simple approximation. For a two-dimensional helium
film at the saturation density, $\sigma$ is about $2.7${\AA} from
numerical calculations.~\cite{Clements93} The corresponding energy
scale associated with this length scale is $\hbar^2/m\sigma^2\approx
1.7K$.

The phase factor in the one-particle integrals in Eqs.
(\ref{eq:kers1}) and (\ref{eq:kers2}) can be rewritten in terms of the
opening angles between the two generator coordinates, $\vr_0$ and
$\vrp_0$,
\begin{equation} 
\phi-\phip=\op+\opi \;,
\end{equation} 
where $\op$ is the opening angle spanned from $\vr-\vrp_0$ to
$\vr-\vr_0$, and $\opi$ is the corresponding angle with respect to the
image vortices. In this form, one can easily see that the integrand is
independent of the coordinate system. These angles can be easily
formulated by considering the coordinates as complex variables,
\begin{eqnarray}
e^{i\op} & = & \sqrt{\frac{(re^{i\theta}-r_0e^{i\theta_0})(re^{-i\theta}-\rp_0e^{-i\thep_0})}{(re^{i\theta}-\rp_0e^{i\thep_0})(re^{-i\theta}-r_0e^{-i\theta_0})}} \;,\label{eq:op}\\
e^{i\opi} & = & \sqrt{\frac{(r\rp_0e^{i\theta}-R^2e^{i\thep_0})(rr_0e^{-i\theta}-R^2e^{-i\theta_0})}{(rr_0e^{i\theta}-R^2e^{i\theta_0})(r\rp_0e^{-i\theta}-R^2e^{-i\thep_0})}} \;.\label{eq:opi}
\end{eqnarray}
One begins the calculation with the assumption that the system size
$R$ is sufficiently large so that $r_0/R$ can be treated as a small
parameter. This assumption is self-consistent only if the weighting
functions are localized far away from the boundary. The validity of
this assumption can be checked later as we discuss the solutions. The
common overlap integral in both $J_l$ and $K_l$ is evaluated by
dividing the integration range into three distinct regions. The first
region is the circle centered at $\vr_c=(\vr_0+\vrp_0)/2$ with
diameter $d=|\vr_0-\vrp_0|$, and has an area $A_1=\pi d^2/4$. In this
region, one can regard the phase contributed by the image vortices as
a constant,
\begin{equation}
\int_0^{2\pi}\frac{d\theta}{2\pi}e^{i\opi}\approx e^{i\Deth_0} \;.
\end{equation}
The remaining part of the integral in the large $R$ limit is
\begin{equation}
\int_{A_1}d^2\vr\,e^{i\op}=A_1-(1.6639+\pi)\left(\frac{d}{2}\right)^2 \;. \label{eq:oia1}
\end{equation}
Since the integrand is dimensionless, the integral must scale as
$d^2$, and the numerical coefficient is calculated explicitly in
Appendix B. The second region is the ring
centered at $\vr_c$ from the inner radius $d/2$ to the outer radius
$R-r_c$, and has an area $A_2=\pi(R-r_c)^2-A_1$. In this region, it is
convenient to shift the origin of my coordinate system to $\vr_c$. The
variables in Eqs.~(\ref{eq:op}) and (\ref{eq:opi}) will be changed
accordingly. I can expand $e^{i\op}$ with respect to $(r_0/r)$ and
$(\rp_0/r)$, and expand $e^{i\opi}$ with respect to $(rr_0/R^2)$ and
$(r\rp_0/R^2)$ up to the second order. Keeping only the terms, which
survive after the angular integration, I obtain
\begin{equation}
\int_0^{2\pi}\frac{d\theta}{2\pi}e^{i(\op+\opi)}\approx e^{i\Deth_0}\left[1-\frac{d^2}{4r^2}\left(1+\frac{r^2}{R^2}\right)^2\right] \;.
\end{equation}
For the radial integral, the upper bound is taken to be $R$, the lower
bound is $d/2$, and the integral gives
\begin{equation}
\int_{A_2}d^2\vr\,e^{i(\op+\opi-\Deth_0)}=A_2-\left(2\ln\frac{2R}{d}+\frac{5}{2}\right)\pi\left(\frac{d}{2}\right)^2 \;, \label{eq:oia2}
\end{equation}
where the factor $5/2$ is contributed by the image vortices. The last
region is simply the remaining area $A_3=A-A_1-A_2$ like a thin
crescent moon. In this region, one can again regard the phase
contributed by the images as a constant, and the remaining integral
gives
\begin{eqnarray}
\int_{A_3}d^2\vr\, e^{i\op} & \approx & \int_{A_3}d^2\vr(1+i\sin\op) \nonumber\\
& = & A_3+i\hat z\cdot\int_{A}d^2\vr\frac{(\vr-\vrp_0)\times(\vr-\vr_0)}{R^2} \nonumber\\
& = & A_3-i\pi r_0\rp_0\sin\Deth_0 \;. \label{eq:oia3}
\end{eqnarray}
In the second line of the above equation, I have changed the
integration range from $A_3$ to $A$ so that the cross terms linear in
$\vr$ vanish. After putting Eqs.~(\ref{eq:oia1}), (\ref{eq:oia2}) and
(\ref{eq:oia3}) together, the original integral is obtained,
\begin{eqnarray}
\lefteqn{ \frac{e^{-i\Deth_0}}{A}\int_A d^2\vr\, e^{i(\op+\opi)} } \nonumber\\
& = & 1-\left(\frac{1}{2}\ln\frac{2R}{d}+\alpha\right)\bigg(\frac{d}{R}\bigg)^2-\frac{ir_0\rp_0\sin\Deth_0}{R^2} \;,
\end{eqnarray}
where $\alpha=7/8+1.6639/4\pi=1.0074$. The overall phase factor
contributed by the image vortices is closely related to the fact that
$l\hbar$ is the total angular momentum of the system, not the angular
momentum of the effective motion of the vortex. This connection will
become clear as we discuss the solutions in the next section.

The other overlap integral in Eq.~(\ref{eq:kers2}) is weighted by the
kinetic energy. The kinetic energy term can also be rewritten in terms
of the opening angles,
\begin{equation}
\nabla\phip\cdot\nabla\phi\simeq\frac{\cos\op}{|\vr-\vr_0||\vr-\vrp_0|} \;.
\end{equation}
In this case the image vortices only contribute to the overall phase
factor. I can again shift the origin to $\vr_c$, and the first
approximation to the integral is
\begin{equation}
\int_0^{2\pi}d\theta\int_{d/2}^R\frac{dr}{r^2}=2\pi\ln(2R/d) \;,
\end{equation}
where $e^{i\op}\approx\cos\op\approx 1$. The correction to this
result must be a dimensionless number since there is no other
dimensional parameters other than $d$,
\begin{equation}
\int_A d^2\vr\nabla\phip\nabla\phi\,e^{i\op}=2\pi\left(\ln\frac{2R}{d}+\beta\right) \;, \label{eq:kin}
\end{equation}
where $\beta=-0.30685$. The detailed calculation of this number can be
found in Appendix B by evaluating the integral numerically. Therefore,
we reach the final form of the kernel in the large $R$ limit as,
\begin{eqnarray}
J_l(\rp_0,r_0) & = & r_0\int_{-\pi}^\pi\!\! d\Deth_0S_l(\rp_0,r_0,\Deth_0) \;,\label{eq:kerf1}\\
K_l(\rp_0,r_0) & = & \frac{\hbar^2}{m}r_0\!\int_{-\pi}^\pi\!\! d\Deth_0\!\left(\ln\frac{2R}{d}+\beta\right)\! S_l(\rp_0,r_0,\Deth_0) \;,\nonumber\\ \label{eq:kerf2}
\end{eqnarray}
where
\begin{eqnarray}
S_l(\rp_0,r_0,\Deth_0) & = & \exp\{-i[r_0\rp_0\sin\Deth_0-(N-l)\Deth_0]\} \nonumber\\
&& \times\exp\left[-\left(\frac{1}{2}\ln\frac{2R}{d}+\alpha\right)d^2\right] \;. \label{eq:kerf3}
\end{eqnarray} 
This kernel is sharply peaked when the two generator coordinates are
close to one another because both the separation $d$ and the phase of
Eq.~(\ref{eq:kerf3}) are small. Notice a peculiar feature that the
kernel depends on the system size explicitly.

\section{Solutions of the integral equation}
\label{sol}

Even though I have greatly simplified the system, the kernel still
remains in an integral form. Neither the analytic form of the kernel,
nor the series solutions of the integral equation have been worked out
because the last integral contains logarithms.  Nevertheless, the
integral equation can be solved numerically, which leads to some
interesting results. In fact, solving this homogeneous integral
equation is basically like solving the eigen-system of a matrix. The
variational energy $\varepsilon_{n,l}$ and the weighting function
$f_{n,l}$ are the eigenvalues and eigenvectors of the matrix
$J_l^{-1}K_l$. The generator coordinates, $r_0$ and $\rp_0$, now only
take discrete values, and act like the matrix indices. The only
complication is that each matrix element of $J_l$ or $K_l$ is a
one-variable integral.

Instead of working with the full-size matrix, in which $r_0$ is
running from $0$ to $R$, it is sufficient to work only with a
sub-matrix, in which $r_0$ is running only from $0$ to a smaller
cut-off $R_{\rm c}$, for two reasons: (a) The matrix elements far away
from the diagonal are considerably smaller than the elements around
the diagonal. This statement is equivalent to say that two far apart
Feynman wave functions are essentially orthogonal to each other. (b)
Since the kernel is sharply peaked, it is reasonable to assume that
the weighting functions are localized in the region where the kernel
is peaked. The use of such cut-off is self-consistent for the
eigenvectors, which are localized in the region bounded by the
cut-off. This self-consistent condition is the same as the one for
calculating the kernel Eqs.~(\ref{eq:kerf1}) and (\ref{eq:kerf2}). The
greatest advantage of this cut-off is that one can maintain the same
numerical accuracy for calculations on fairly large system sizes
without drastically increasing the sizes of the matrices. A typical
matrix in my calculation has about two hundred elements on each side
for $R_{\rm c}=10$. The calculation is carried out for a series of
different sizes of matrices at a fixed $R_{\rm c}$, and an
extrapolation to the continuum limit is performed in the end to obtain
the final results.

\begin{figure}
\begin{tabular}{cc}
\epsfxsize 3.4in
\epsffile{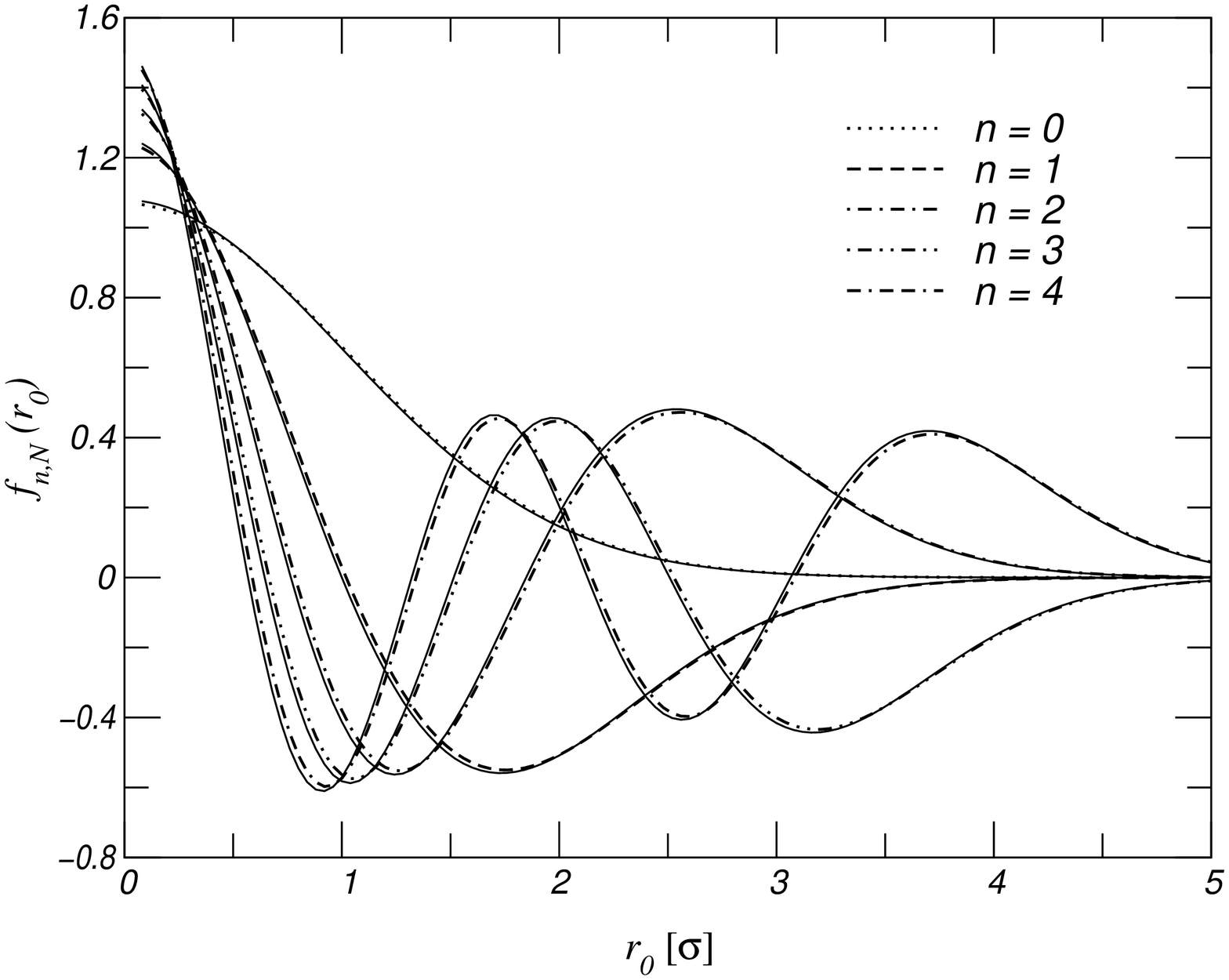} &
\epsfxsize 3.4in
\epsffile{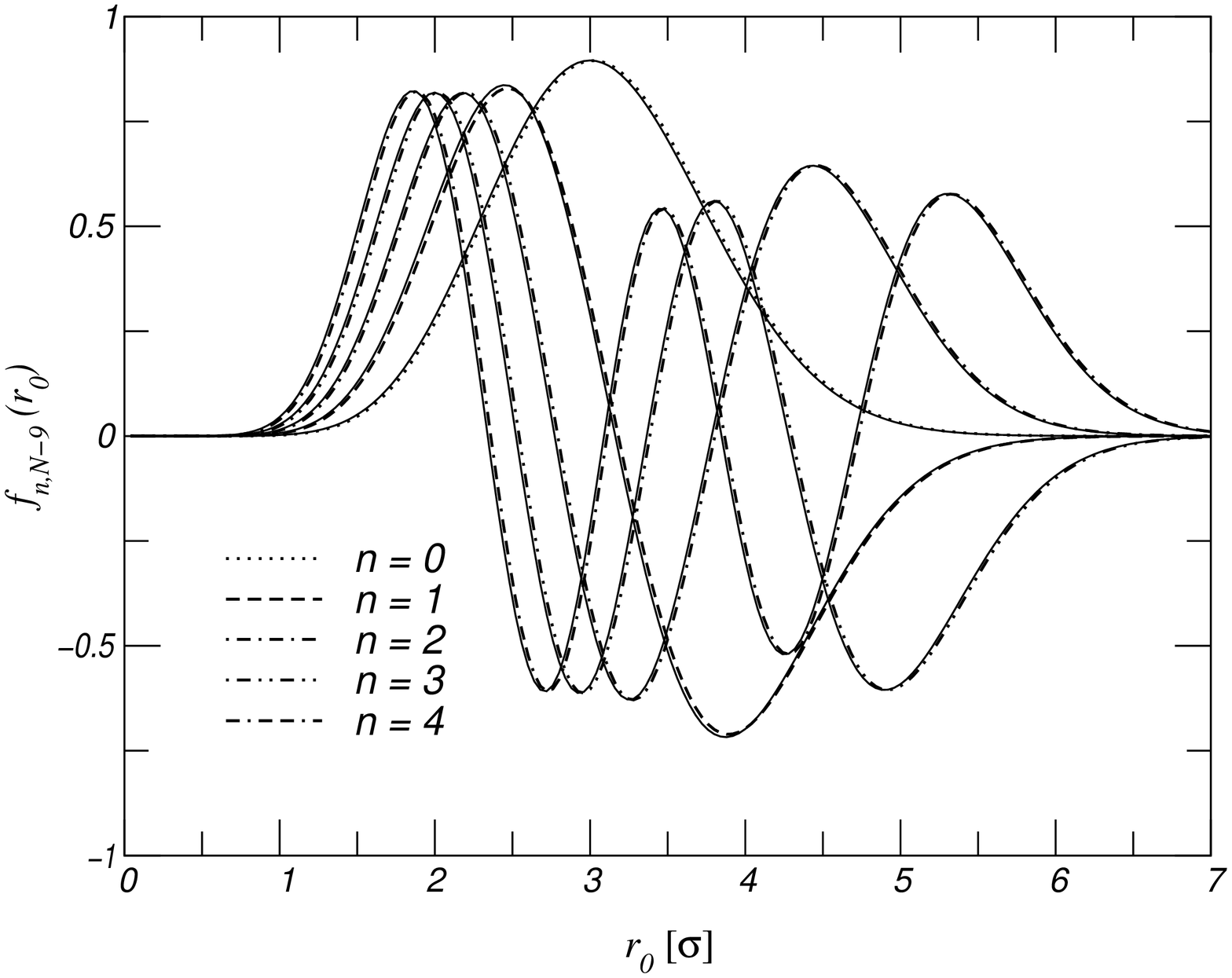} \\
\epsfxsize 3.4in
\epsffile{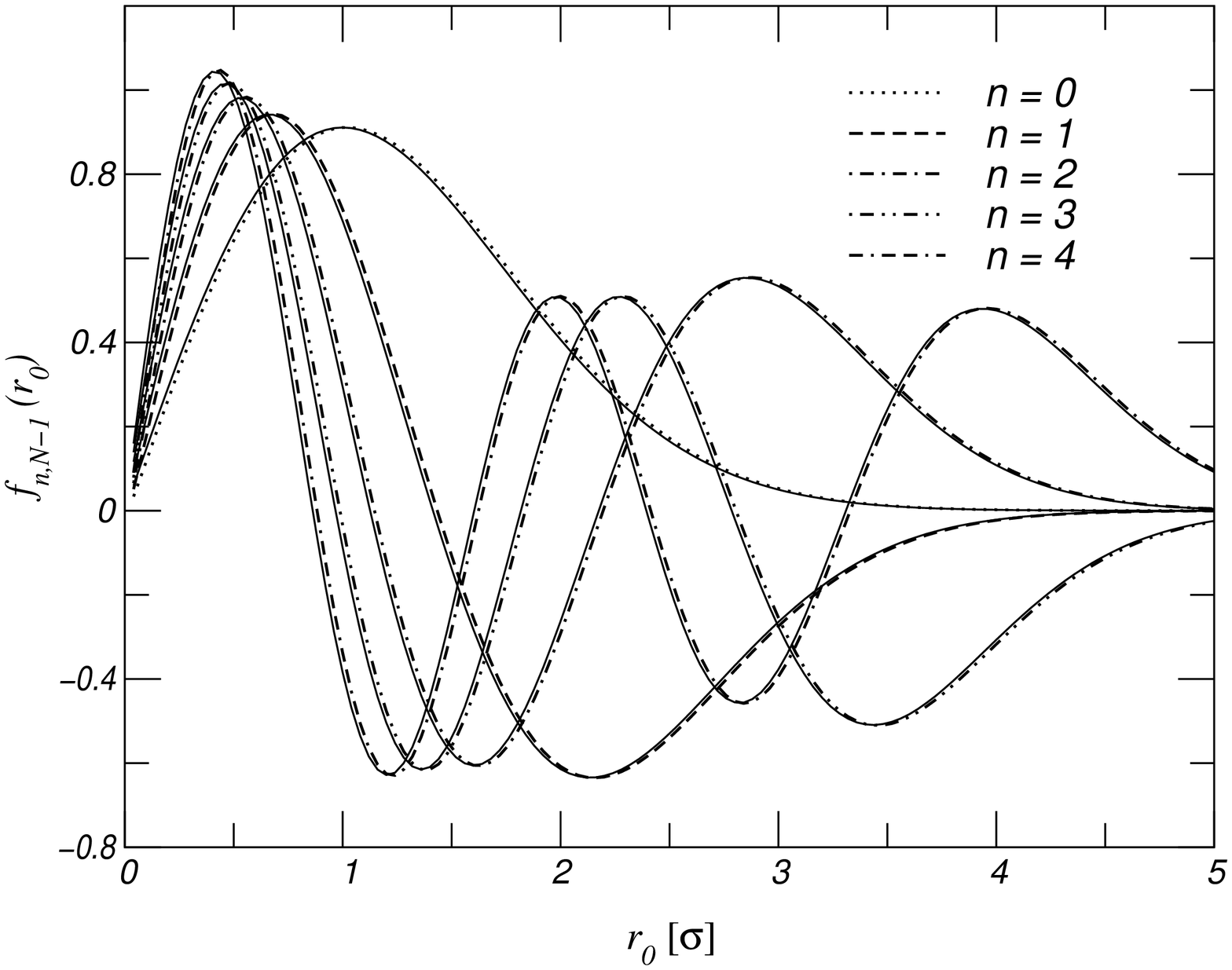} &
\epsfxsize 3.4in
\epsffile{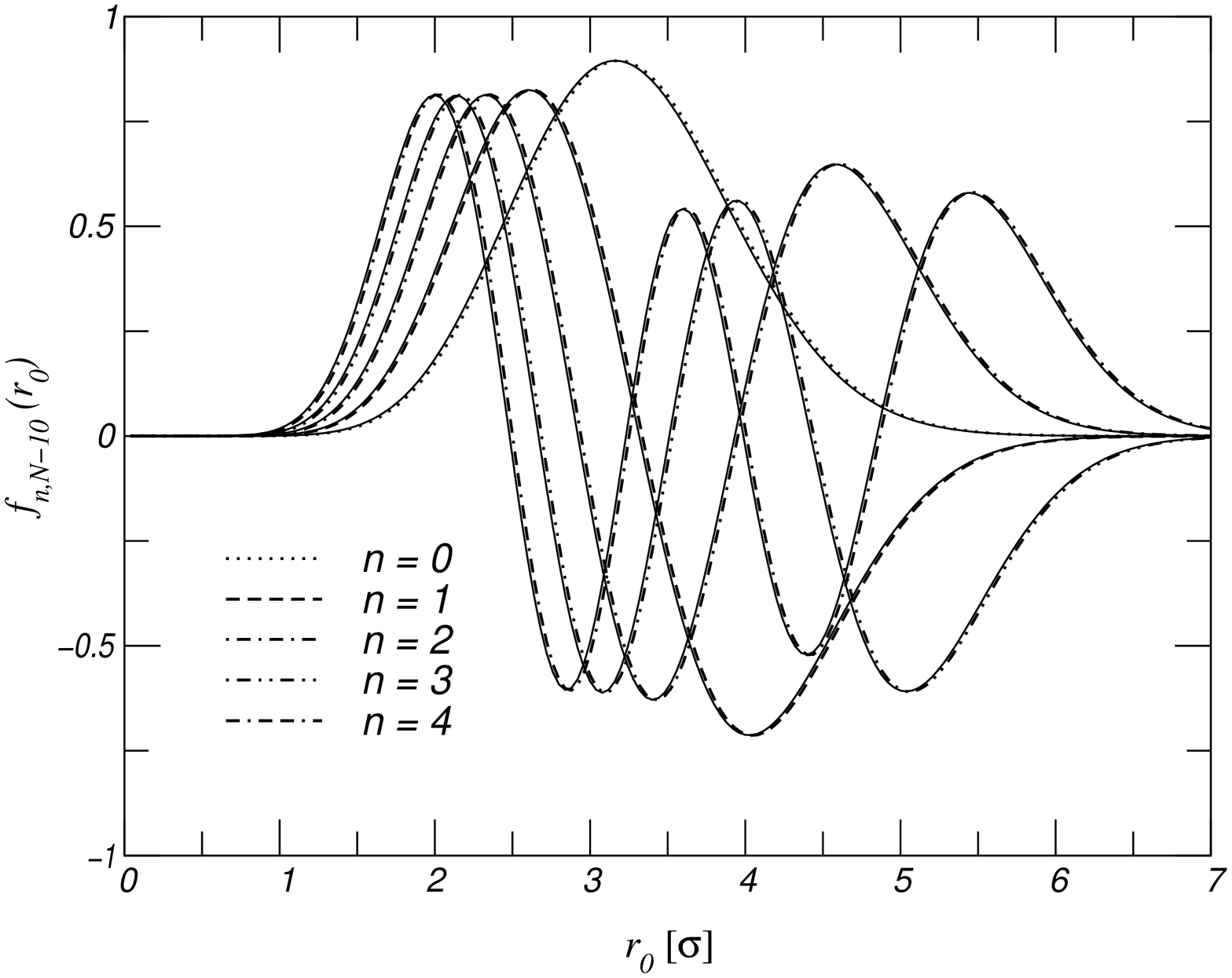}
\end{tabular}
\widetext
\caption{ The normalized weighting functions for angular momenta
$l=N,N-1,N-9,N-10$. These are the eigenfunctions of matrices with
$250\times 250$ elements at the system size $R=10$. The normalization
condition is $\int drf_{n,l}(r)^2=1$. The solid lines are the
corresponding radial wave functions of an electron. The only
adjustable parameter for the solid curves is the normalization
constant. }
\label{fig:wf10}
\end{figure}

\begin{figure}
\epsfxsize \columnwidth
\epsffile{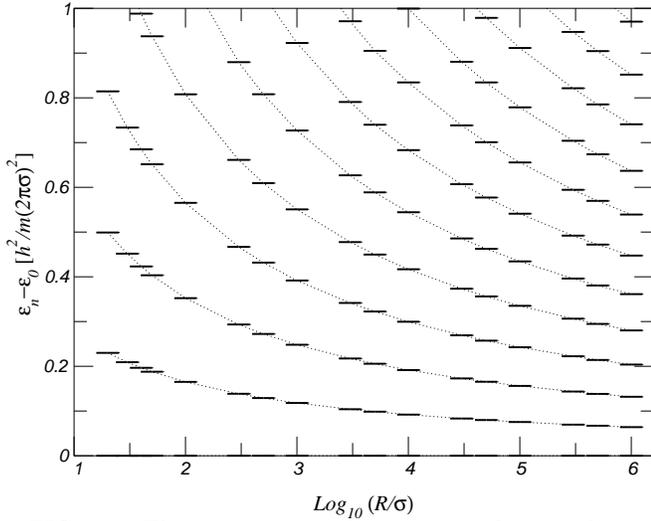}
\caption{ The energy spectrum corresponding to the cyclotron motion
of a vortex relative to the lowest energy level. The dotted lines show
the scaling of each energy level with the system radius $R$ ranging
from $20\sigma$ to $10^6\sigma$. The lowest energy level approximately
scales as $\ln(R/\sigma)$ corresponding to the formation energy of a
vortex. The same spectrum is obtained for any given angular
momentum. }
\label{fig:engsp}
\end{figure}

Numerical results are summarized in Fig.~\ref{fig:wf10} and
Fig.~\ref{fig:engsp}. Fig.~\ref{fig:wf10} represents a few examples of
the weighting functions, whose modulus squares are normalized to
unity. These weighting functions are well-localized, so that the
self-consistent condition is justified. Their functional forms are
found to be independent of the system size as long as $R$ is large
enough to satisfy the self-consistent condition. In general, the
weighting functions should depend both on the system size and the
angular momentum, because the kernel depends on both parameters in a
non-trivial way. Besides the evidence from numerical solutions, I am
unable to provide any analytical calculation here to prove this
independence rigorously. Physically, this independence means that the
motion of a vortex is not sensitive to the boundary conditions, which
is quit reasonable. Fig.~\ref{fig:engsp} is the energy spectrum for
various different system sizes. The angular momentum index is dropped
because the energy eigenvalues in the continuum limit are independent
of the angular momentum. These large degeneracies in the variational
energies can be understood approximately, in the large $R$ limit, as
follows. For clarity, I will only show the equations for $J_l$ because
the equations for $K_l$ are almost identical. For weighting functions
sufficiently far away from the origin, the kernel of their integral
equation is peaked in a small region around
$r_0\approx\rp_0\approx\sqrt{N-l}$ with $|N-l|\gg 1$. The kernel can
be approximated by
\begin{equation}
J_l\approx\int_{-\infty}^\infty dye^{-(\ln\sqrt{2R/d}+\alpha)d^2}e^{-iy[\rp_0-(N-l)/\rp_0]} \;, \label{eq:kerap}
\end{equation}
where the variables are changed to $x=|\rp_0-r_0|$, $y=r_0\Deth_0$ and
$d^2\approx x^2+y^2$. Two kernels with different angular momenta are
related to each other in the following way,
\begin{equation}
J_{l+\Delta l}(\rp_0+\Delta r_l,r_0)\approx J_l(\rp_0,r_0) \;.
\end{equation}
For a given $\Delta l$, one can find that $\Delta r_l=\sqrt{N-l-\Delta
l}-\sqrt{N-l}$ by keeping the phase of the integrand in
Eq.~(\ref{eq:kerap}) invariant when $\rp_0$ is near
$\sqrt{N-l}$. Therefore, there are approximate relations among the
variational energies and among the weighting functions,
\begin{eqnarray}
\varepsilon_{n,l+\Delta l} & \approx & \varepsilon_{n,l} \;,\\
f_{n,l+\Delta l}(r_0+\Delta r_0) & \approx & f_{n,l}(r_0) \;.
\end{eqnarray}
As a result, the variational energies are highly degenerate, and the
degeneracies are of order $N$. The similarity between any two
weighting functions at the same energy level with different angular
momenta is demonstrated by comparing the plots with $l=N-9$ and with
$l=N-10$ in Fig.~\ref{fig:wf10}.

\begin{figure}
\epsfxsize \columnwidth
\epsffile{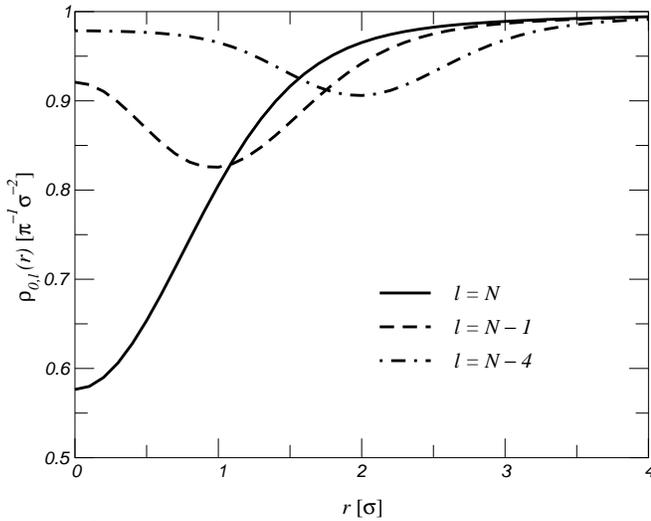}
\caption{ The radial profiles of the number density for three different
states, each of which has total angular momentum $N$, $N-1$, and
$N-4$, in the first Landau level. The curves represent the numerical
results in a finite system with radius $R=10$. The limit for all
states in the infinite system is uniform density. Instead of fixing
the total number of particles, the normalization is chosen in such a
way that the asymptotic value is unity. }
\label{fig:dens}
\end{figure}

To see the structure of a vortex in this model, one can compute the
number density and the current density. The number density is
\begin{equation}
\rho_{n,l}(\vr)=\dnvr\Psi_{n,l}^*\left[\sum_{j=1}^N\delta(\vr-\vr_j)\right]\Psi_{n,l} \;.
\end{equation}
Since the density is independent of the azimuthal angle by symmetry, I
can perform an angular average, and rewrite the integrand as a
function of $\Deth_0$. The radial profile of the density is
\begin{eqnarray}
\rho_{n,l}(r) & = & \frac{N}{A}\int_{-\pi}^\pi d\Deth_0\weighting \nonumber\\
&& \times S_l(\rp_0,r_0,\Deth_0)\int_0^{2\pi}d\theta\,e^{i(\phi-\phip-\Deth_0)} \;.\label{eq:density}
\end{eqnarray}
The angular average of the last integral in the above formula has been
worked out explicitly as elliptic integrals in the appendix,
Eq.~(\ref{eq:ang1}). A few examples for vortex states in the lowest
energy level, $n=0$, are shown in Fig.~\ref{fig:dens}. The density
somewhat drops near the core, but generally speaking is quite
uniform. The density never drops to zero at the center of the
vortex. All the density profiles depend logarithmically on the system
size, and become completely uniform in the infinite system limit. This
result is quite consistent with the conjecture that the density is
roughly constant even inside the vortex core, proposed by
Fetter~\cite{Fetter71} based on the argument of strong pair
correlations in the wave function. In comparison with the Monte Carlo
calculations for a static vortex,~\cite{Ortiz95,Vitiello96,Sadd97} my
density profiles look like ``spatially averaged'', and miss several
detailed features on the scale of particle spacing. For examples, the
fractional density at the core center is over-estimated because my
cut-off function $g(r)$ is set to unity. Spatial oscillations due to
higher-order correlation in the ground state are neglected by the
Hartree approximation. The core size is also over-estimated possibly
because the strong repulsion among particles has been under estimated
in the dilute limit. In liquid helium, the range, in which the
two-body potential is strongly repulsive, is comparable to the spacing
between two helium atoms.

\begin{figure}
\epsfxsize \columnwidth
\epsffile{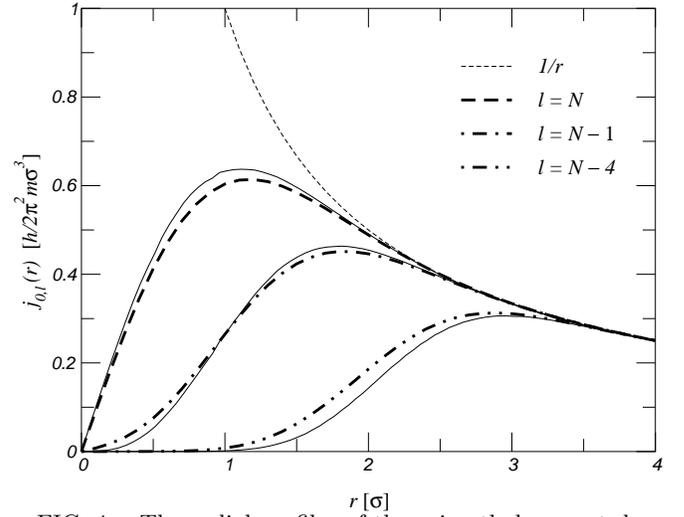}
\caption{ The radial profiles of the azimuthal current density for
three different states, each of which has total angular momentum $N$,
$N-1$, and $N-4$, in the first Landau level. The dashed lines
represent the numerical results in a finite system with radius $R=10$,
and the solid lines represent the results for the infinite-system
limit. The dotted line is the classical result for a singular vortex
line. }
\label{fig:velo}
\end{figure}

The current density in the azimuthal direction is $j_{n,l}(\vr)$,
\begin{equation}
\Im\left\{\frac{\hbar}{m}\dnvr\Psi_{n,l}^*\left[\sum_{j=1}^N\delta(\vr-\vr_j)\frac{1}{r}\frac{\partial}{\partial\theta}\right]\Psi_{n,l}\right\} \;.
\end{equation}
Once again, I can make an angular average so that the integrand
depends only on $\Deth_0$,
\begin{eqnarray}
\frac{N}{A}\frac{\hbar}{mr}\int_{-\pi}^\pi d\Deth_0\weighting \nonumber\\
\times\Im\left[S_l(\rp_0,r_0,\Deth_0)e^{-i\Deth_0}\int_0^{2\pi}d\theta\,e^{-i\phip}\frac{\partial}{\partial\theta}e^{i\phi}\right] \,.\label{eq:current}
\end{eqnarray}
The analytic form of the last angular integral has been worked out in
terms of elliptic integrals in the appendix, Eq.~(\ref{eq:ang3}). A
few plots of the azimuthal current density for vortex states in the
lowest energy level are shown in Fig.~\ref{fig:velo}. The current
density is always zero at the origin as expected, and behaves like
$1/r$ asymptotically. All current distributions are approximately
independent of the system size. The vorticity is distributed around
the vortex core over a region for about two or three atomic layers.

\section{Landau Levels and Effective Mass}
\label{mass}

The motion of a vortex can now be understood by comparing the
weighting functions in Fig.~\ref{fig:wf10}
to the wave functions of an electron in an uniform magnetic field $B$.
In the central gauge, $\vA=(-yB/2,xB/2,0)$, the Hamiltonian for the
electron is
\begin{equation}
H=\frac{(\vp-e\vA)^2}{2M}
=\frac{{\vp}^2}{2M}+\frac{M}{2}\left(\frac{\omega}{2}\right)^2r^2-s\frac{\omega}{2}L_z  \;, \label{eq:electron}
\end{equation}
where $\omega=|eB|/M$ is the cyclotron frequency, $L_z=xp_y-yp_x$ is
the $z$ component of the angular momentum operator, and $s$ is the
sign of $eB$. Without the $L_z$ term, the Hamiltonian represents a
two-dimensional simple-harmonic oscillator, which has energy
eigenvalues $(n+1)\hbar\omega/2$, where $n$ is a non-negative
integer. The characteristic frequency for this oscillator is only half
of the cyclotron frequency. The degeneracy for the $n$-th level is
$n+1$. With the $L_z$ term, all energies with odd(even) $n$ are
shifted by an odd(even) multiples of $\hbar\omega/2$, and the energy
spectrum changes to $E=(n+1/2)\hbar\omega$. The degeneracies for the
low-lying levels are now proportional to the system size. These energy
levels are called Landau levels. In the polar coordinates, the radial
wave function $R_{n,m}(r)$ in the $n$-th Landau level with angular
momentum $m\hbar$ is a confluent hypergeometric function shown in
Appendix A. There is also only one length scale set by the strength of
the magnetic field. If this length scale, $\sqrt{2\hbar/|eB|}$, is
chosen to be the unit length for the electron problem, the weighting
function $f_{n,N-m}(r_0)$ can be matched perfectly with the radial
wave function $R_{n,sm}(r)$. This one-to-one correspondence
immediately tells us that the motion of a vortex is indeed
cyclotron-like, and the degeneracy of each energy level in
Fig.~\ref{fig:engsp} is one per $\pi$ unit area, since the degeneracy
of each Landau level is $|eB|/h$ per unit area.

Once the basis for the cyclotron motion of a vortex is established,
the parameter, namely the effective mass, in the phenomenological
theory can be derived from this microscopic theory. In cyclotron
motion, the effective mass is related to the inverse of the
energy-level spacing,
\begin{equation}
M_{\rm eff}=\frac{\hbar\rho\kappa}{\DE}=\frac{2\hbar^2}{\sigma^2\DE} \;,
\end{equation}
where $\DE$ is the energy-level spacing, and the role of $eB$ is
replaced by $\rho\kappa$. $\rho$ is the asymptotic mass density of the
fluid, $m/\pi\sigma^2$, and $\kappa$ is the quantum of circulation,
$h/m$. However, the energy spectrum in Fig.~\ref{fig:engsp} is
slightly deviated from the simple-harmonic spectrum that the level
spacings tend to increase with energy. Each energy level, therefore,
has a different effective mass. At present, it is not clear whether
this deviation is due to the deficiency of the projection method, or
is a real many-body effect. The single-particle picture is in fact
broken by this small deviation because one cannot simply add an
attractive potential in Eq.~(\ref{eq:electron}) to mimic the spectrum.
The degeneracy of the Landau level will be broken in the same time by
the additional potential. On the other hand, it is known, in the case
of a nucleus, that the projection method in its simplest version only
gives an approximate answer for the mass. It was pointed out by
Peierls and Thouless~\cite{Peierls62} that the mass of a nucleus is
only given exactly by working with further improved wave functions,
which allow quantum fluctuations not only in the position but also in
the velocity. In the case of a vortex, their suggestion is equivalent
to taking the backflow into account, and considering trial wave
functions of the following form,
\begin{equation}
\Psi=\int d^2\vr_0d^2\vv_0G(\vr_0,\vv_0)\Psi_{\vr_0,\vv_0} \;.
\end{equation}
The fluctuation in velocities is usually separable from the
fluctuation in positions, and takes a simple Gaussian form. The effect
of the backflow is not clear at the moment, and is currently under
investigation.

\begin{figure}
\epsfxsize \columnwidth
\epsffile{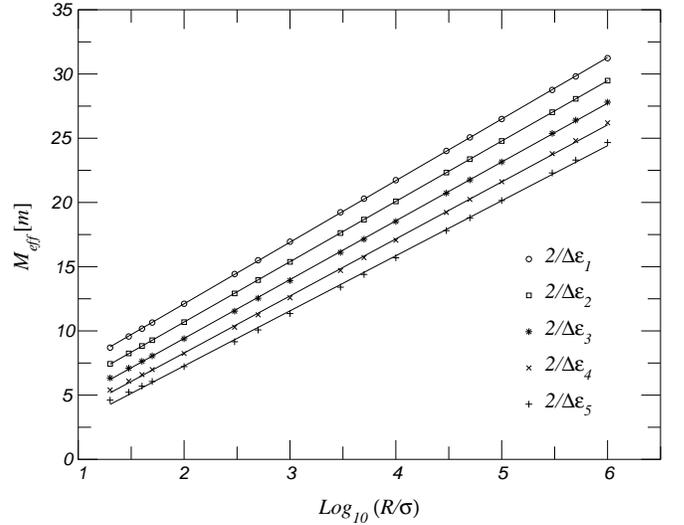}
\caption{ The scalings of the inverse of energy-level spacings labeled
as $\DE_n=\varepsilon_n-\varepsilon_{n-1}$, which are related to the
effective mass of a vortex. The symbols show the first five of them
from top to bottom. The lines are the least-square fits, and clearly
show that the effective mass scales logarithmically with the vortex
size $R$. }
\label{fig:mass}
\end{figure}

The effective mass of a quantized vortex is a controversial subject
even qualitatively.~\cite{Duan94,Duan95,Niu94,Niu95} One important
issue is whether the mass scales with the system size. I have
calculated the effective masses of the first few energy levels for
various system sizes, and plot them in Fig.~\ref{fig:mass}. The data
clearly shows that the effective mass scales logarithmically with the
size of the vortex. This qualitative feature is not affected by the
issue discussed in the end of the previous paragraph because the
correction from the backflow is clearly finite. This special scaling
behavior is attributed to the long-ranged phase coherence in a
superfluid, which causes the logarithmic size-dependence of the
overlap in Eqs.~(\ref{eq:kerf1}) to (\ref{eq:kerf3}) between any two
Feynman wave functions. It is not surprising that the overlap is
significant non-zero only when the separation of two vortex center
coordinates are close to each other. However, the functional form of
the kernel for a short-ranged correlated system such as a nucleus is
quite different because the width of the peak does not scale
logarithmically with the system size.  Even though the calculation is
done rigorously only for a system of weakly interacting bosons, I now
argue that the scaling behavior is going to persist even in a strongly
interacting system. The orthogonality is due to the phase factor
associated with the vortex. Any two shifted vortex wave functions
eventually lose their phase coherence at the large distance, even
though they are only displaced by a small amount. Interaction may
change the detail dependence on the vortex coordinates or the
constants, but can only change the short-distance physics. For a
sufficiently large vortex, the logarithm is going to be the dominate
term, so is the scaling behavior. In a real system, the logarithm must
be cut off at a certain length scale, which could be the distance that
the condensate loses its phase coherence, or the spacing among
vortices.

In conclusion, I have presented a microscopic quantum theory to
support the picture that a quantized vortex behaves like an electron
in a magnetic field. Vortex states with different angular momenta form
highly degenerate levels. The density is finite at the vortex core
axis, and the vorticity is distributed inside the core. The dynamical
effective mass defined as the inverse of the energy-level spacing
scales logarithmically with the size of the vortex.

\acknowledgments

I am very grateful to David Thouless for initiating the idea of this
work and stimulating discussions. I also like to thank Ping Ao for
valuable comments and suggestions. This work was partially supported
by National Science Foundation under Grant No. DMR-9528345 and
DMR-9815932.

\onecolumn
\appendix

\section{An Electron in a Magnetic Field}

In this appendix I show the energy eigenfunctions of an electron in a
uniform magnetic field.~\cite{Landau77} I look for energy
eigenfunctions in the polar coordinates, $R_{n,m}(r)e^{im\theta}$. The
radial Schr\"odinger equation takes the following form,
\begin{equation}
\frac{\hbar^2}{2M}\left[\frac{\partial^2}{\partial r^2}+\frac{1}{r}\frac{\partial}{\partial r}+\frac{2ME}{\hbar^2}-\left(\frac{m}{r}-\frac{M\omega r}{2\hbar}\right)^2\right]R_{n,m}(r)=0 \;,
\end{equation}
where $E=(n+1/2)\hbar\omega$. It is obvious from the differential
equation that the wave function falls off exponentially in the
asymptotic region. In dimensionless variable
$\xi=r/\sqrt{2\hbar/|eB|}$, the radial wave function takes the
following form,
\begin{equation}
R_{n,m}(\xi)=g_{n,m}(\xi)e^{-\xi^2/2} \;,
\end{equation}
where $g_{n,m}(\xi)$ satisfies the differential equation,
\begin{equation}
\left[\frac{\partial^2}{\partial\xi^2}+\left(\frac{1}{\xi}-2\xi\right)\frac{\partial}{\partial\xi}+2(2n+sm)-\frac{m^2}{\xi^2}\right]g_{n,m}(\xi)=0 \;.
\end{equation}
In looking for the series solutions of $g_{n,m}(\xi)$, the indicial
equation has two roots, $\pm m$. Since the wave function is required
to be regular at the origin, only the positive root satisfies the
requirement,
\begin{equation}
g_{n,m}(\xi)=\xi^{|m|}\sum_{j=0}^\infty c_j\xi^j \;,
\end{equation}
where $c_j$ satisfies the following recursion relation,
\begin{equation}
c_{j+2}=2\left[\frac{|m|+j-2n-sm}{(|m|+j+2)^2-m^2}\right]c_j \;.
\end{equation}
The energy is, therefore, quantized so that the infinite series
terminates at the finite order $j_{\rm max}=2n+sm-|m|$. This series
can be rewritten in terms of the confluent hypergeometric function,
\begin{equation}
g_{n,m}(\xi)=\xi^{|m|}\,_1\!F_1\left[-(2n+sm-|m|)/2,|m|+1;\xi^2\right] \;,
\end{equation}
where $_1\!F_1(a,b;x)$ is defined as
\begin{equation}
_1\!F_1(a,b;x)=\sum_{j=0}^\infty\frac{(a+j-1)!}{(b+j-1)!}\frac{x^j}{j!} \;.
\end{equation}
Apart from the normalization constant, $R_{n,sm}(\xi)$ is
identical to the weighting function $f_{n,N-m}(r_0)$
in Eq.~(\ref{eq:trial}) as shown in Fig.~\ref{fig:wf10}.

\section{Angular Integrals}

This appendix contains several angular integrals for evaluating the
constants in Eqs.~(\ref{eq:oia1}) and (\ref{eq:kin}), and for
calculating the radial profile of the number density in
Eq.~(\ref{eq:density}) and the current density in
Eq.~(\ref{eq:current}). The remaining radial parts of these integrals
can then be carried out numerically in a reasonably length of time.

\begin{figure}
\epsfxsize 0.75\columnwidth
\centerline{\epsffile{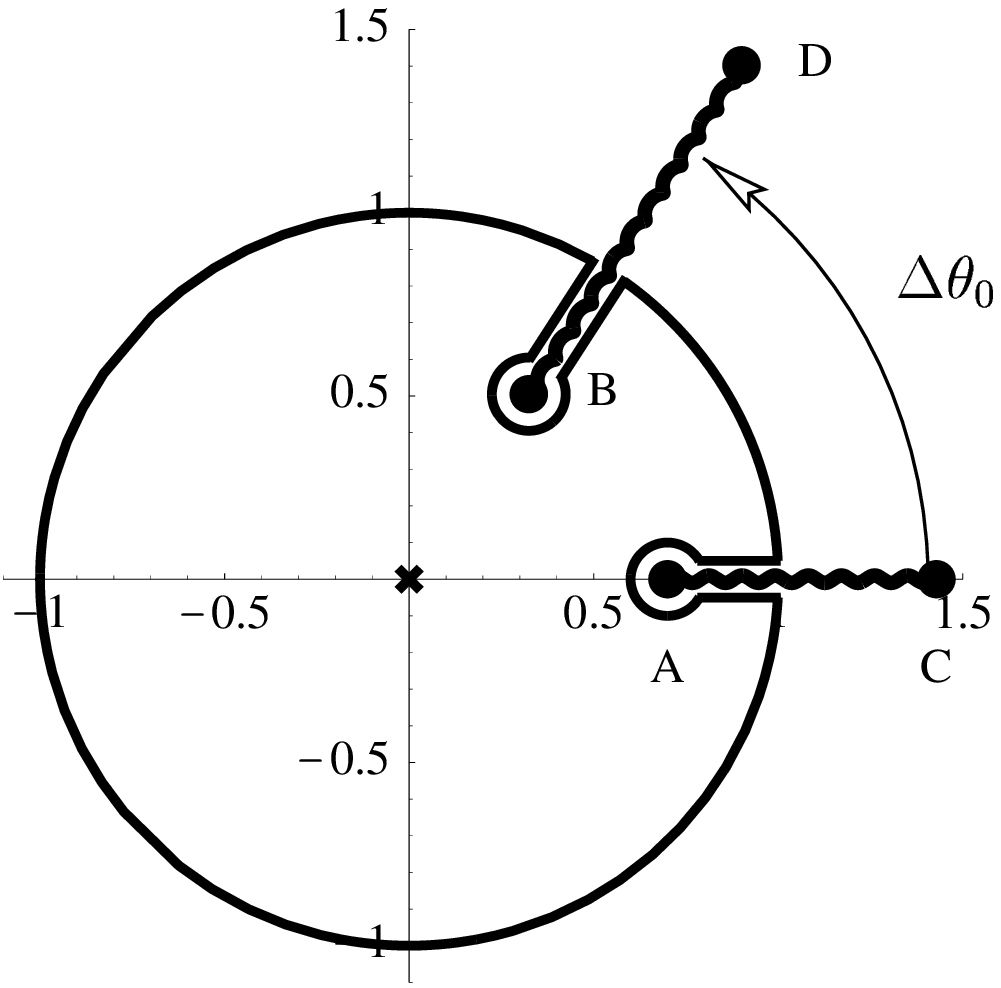}}
\caption{ Branch points, $A$ to $D$, branch cuts, $\overline{AC}$
and $\overline{BD}$, and the integration contour deformed from the
unit circle. This particular sequence of the branch points corresponds
to the case $r>r_0,\rp_0$. There may be a simple pole located at the
origin depending on which integrand is discussed. }
\label{fig:contour}
\end{figure}

The overlap integral in Eq.~(\ref{eq:oia1}) can be written in the
following form using complex variables,
\begin{eqnarray} 
\int d^2\vr\, e^{i\op} & = & \int drr\int d\theta\sqrt{\frac{(re^{i\theta}-r_0e^{i\theta_0})(re^{-i\theta}-\rp_0e^{-i\thep_0})}{(re^{i\theta}-\rp_0e^{i\thep_0})(re^{-i\theta}-r_0e^{-i\theta_0})}} \;,\\
& = & \int drr\sqrt{\frac{C}{B}}\oint\frac{dz}{iz}\sqrt{\frac{(z-A)(z-B)}{(z-C)(z-D)}} \;,\label{eq:form1}
\end{eqnarray}
where
\begin{equation}
\begin{array}{cccc}
A=\dfrac{r_0}{r}e^{i\theta_0} & B=\dfrac{r}{\rp_0}e^{i\thep_0} & C=\dfrac{r}{r_0}e^{i\theta_0} & D=\dfrac{\rp_0}{r}e^{i\thep_0}
\end{array} \;.
\end{equation}
On the complex plane, there are two Riemann sheets and four branch
points. The integration contour is along the unit circle, and there
are always two branch points inside the contour and two points outside
the contour. One can always rotate the axes, so that two branch points
are lying on the $x$-axis. I choose the branch cuts in such a way that
the contour is separated into two disjointed curves. The contour and
the branch cuts are shown in Fig.~\ref{fig:contour}. Using the formula
for indefinite integrals obtained in the next appendix, I can carry
out the angular part of the integral by suitably deforming the
contours. Two additional contours are added to go around the branch
points, and to avoid the branch cuts. The integral along the circular
contour is now equal to the sum of four integrals on the straight
segments plus the residue from a simple pole at the origin. The result
is summarized as follows
\begin{equation}
\left\{\begin{array}{l}
2\pi\sqrt{\dfrac{r_0}{\rp_0}}e^{-i\Deth_0/2}+4C_1\Big[\Pi(n_1,k)-\Pi(n_2,k)\Big] \hfill\begin{array}{ll} {\rm for} & r>r_0,\rp_0 \end{array} \\
2\pi\sqrt{\dfrac{r_0}{\rp_0}}e^{-i\Deth_0/2}-4C_1\bigg\{\Big[\Pi(n_1,k)-\Pi(n_2,k)\Big]+\dfrac{1}{\sqrt{k}}\bigg[\Pi\Big(\dfrac{1}{n_3},\frac{1}{k}\Big)-\Pi\Big(\dfrac{1}{n_4},\frac{1}{k}\Big)\bigg]\bigg\} \\
\hfill\begin{array}{ll} {\rm for} & r_0<r<\rp_0 \end{array} \\
2\pi\sqrt{\dfrac{r_0}{\rp_0}}e^{-i\Deth_0/2}-4C_1\Big[\Pi(n_3,k)-\Pi(n_4,k)\Big] \hfill\begin{array}{ll} {\rm for} & r<r_0,\rp_0 \end{array}
\end{array}\right. \;,\label{eq:ang1}
\end{equation}
where $\Pi(n,k)=\Pi(n,\pi/2,k)$ is the complete elliptic integral
defined in the next appendix. Various abbreviations in the equation
are defined as follows
\begin{eqnarray}
C_1 &=& \frac{r^2-r_0\rp_0e^{-i\Deth_0}}{\sqrt{(r^2-r_0^2)(r^2-{\rp_0}^2)}} \;,\\
n_1 &=& \dfrac{\rp_0(r_0e^{-i\Deth_0}-\rp_0)}{r^2-{\rp_0}^2} \;,\\
n_2 &=& \dfrac{r^2(r_0-\rp_0e^{i\Deth_0})}{r_0(r^2-{\rp_0}^2)} \;,\\
k &=& \dfrac{-r^2(r_0^2+{\rp_0}^2-2r_0\rp_0\cos\Deth_0)}{(r^2-r_0^2)(r^2-{\rp_0}^2)} \;,
\end{eqnarray}
and $(n_3, n_4)$ are simply $(n_1, n_2)$ with $(r_0, \rp_0)$
interchanging their roles. This is also the formula for calculating
the number density in Eq.~(\ref{eq:density}).

The overlap integral with the kinetic energy term in
Eq.~(\ref{eq:kin}) can be done in the similar way,
\begin{eqnarray}
\lefteqn{\int d^2\vr\nabla\phip\nabla\phi e^{i\op}} \nonumber\\
& = &\int drr\int d\theta\frac{(re^{i\theta}-r_0e^{i\theta_0})(re^{-i\theta}-\rp_0e^{-i\thep_0})+(re^{i\theta}-\rp_0e^{i\thep_0})(re^{-i\theta}-r_0e^{-i\theta_0})} {2(re^{i\theta}-\rp_0e^{i\thep_0})(re^{-i\theta}-r_0e^{-i\theta_0})\big|re^{i\theta}-r_0e^{i\theta_0}\big|\big|re^{i\theta}-\rp_0e^{i\thep_0}\big|} \\
& = & \int drr\oint\frac{dz}{2ir^2}\left[\frac{\sqrt{BC}}{\sqrt{(z-A)(z-B)(z-C)(z-D)}}+\sqrt{\frac{C^3(z-A)(z-B)}{B(z-C)^3(z-D)^3}}\right] \;.\label{eq:form2}
\end{eqnarray}
Using the same contour without the simple pole at the origin, I obtain
the angular integral as
\begin{equation}
\left\{\begin{array}{l}
4\Big[C_2E(k)-iC_3K(k)\Big] \hfill\begin{array}{lll} {\rm for} & r>r_0,\rp_0 & r<r_0,\rp_0 \end{array} \\
-4C_2\Big[E(k)-\dfrac{(1-k)}{\sqrt{k}}K(1/k)+\sqrt{k}E(1/k)\Big]+4iC_3\bigg[K(k)-\dfrac{K(1/k)}{\sqrt{k}}\bigg] \\
\hfill\begin{array}{ll} {\rm for} & r_0<r<\rp_0 \end{array}
\end{array}\right. \;,\label{eq:ang2}
\end{equation}
where $E(k)=E(\pi/2,k)$ and $K(k)=F(\pi/2,k)$ are again the complete
elliptic integrals defined in the next appendix, and
\begin{eqnarray}
C_2 & = & \frac{\sqrt{(r^2-r_0^2)(r^2-{\rp_0}^2)}}{(r^2-r_0\rp_0e^{i\Deth_0})^2} \;,\\
C_3 & = & \frac{r_0\rp_0\sin\Deth_0}{(r^2-r_0\rp_0e^{i\Deth_0})\sqrt{(r^2-r_0^2)(r^2-{\rp_0}^2)}} \;.
\end{eqnarray}

To calculate the azimuthal current density in Eq.~(\ref{eq:current}),
I have the following integral,
\begin{eqnarray}
\lefteqn{-ie^{-i\Deth_0}\int\frac{d\theta}{2r}\left[e^{-i\phip}\left(\frac{\partial}{\partial\theta}e^{i\phi}\right)-\left(\frac{\partial}{\partial\theta}e^{-i\phip}\right)e^{i\phi}\right]} \\
& = & \sqrt{\frac{C}{B}}\oint\frac{dz}{4irz}\left(\frac{z}{z-A}+\frac{z}{z-D}-\frac{B}{z-B}-\frac{C}{z-C}\right)\sqrt{\frac{(z-A)(z-B)}{(z-C)(z-D)}} \;.\label{eq:form3}
\end{eqnarray}
The result for $r>r_0,\rp_0$ is
\begin{equation}
\frac{\pi}{r}\sqrt{\frac{r_0}{\rp_0}}e^{-i\Deth_0/2}+2C_4\Big[\Pi(n_1,k)-\Pi(n_2,k)\Big]+2C_5E(k) \;, \label{eq:ang3}
\end{equation}
where 
\begin{eqnarray}
C_4 & = & \frac{r^2-r_0\rp_0e^{-i\Deth_0}}{r\sqrt{(r^2-r_0^2)(r^2-{\rp_0}^2)}} \;, \\
C_5 & = & \frac{\sqrt{(r^2-r_0^2)(r^2-{\rp_0}^2)}}{r(r^2-r_0\rp_0e^{i\Deth_0})} \;.
\end{eqnarray}
For $r_0<r<\rp_0$ and $r<r_0,\rp_0$, similar substitutions can be
performed as in the previous cases, Eqs.~(\ref{eq:ang1}) and
(\ref{eq:ang2}).

\section{Formula with Elliptic Integrals}

The following indefinite integrals are the basic ingredients in the
previous appendix. These integral are calculated with the help of {\em
Mathematca}. For the overlap integral in Eq.~(\ref{eq:form1}), I use
\begin{equation}
\int\frac{dz}{z}\sqrt{\dfrac{(z-a)(z-b)}{(z-c)(z-d)}}=\frac{2(a-b)}{\sqrt{(a-c)(b-d)}}\Big[\Pi(n,\phi,k)-\Pi(bn/a,\phi,k)\Big] \;,
\end{equation}
where $\Pi(n,\phi,k)$ is the elliptic integral of the third kind,
\begin{equation}
\Pi(n,\phi,k)=\int_0^\phi\frac{d\theta}{(1-n\sin^2\theta)\sqrt{1-k\sin^2\theta}} \;,
\end{equation}
and the parameters are defined as
\begin{eqnarray}
n & = & \dfrac{a-d}{b-d} \;,\\
\phi & = & \sin^{-1}\sqrt{\dfrac{(b-d)(z-a)}{(a-d)(z-b)}} \;,\\
k & = & \dfrac{(b-c)(a-d)}{(a-c)(b-d)} \;.
\end{eqnarray}
For the overlap integral weighted by the kinetic energy in
Eq.~(\ref{eq:form2}), two indefinite integrals are needed. The first
one is
\begin{equation}
\int\frac{dz}{\sqrt{(z-a)(z-b)(z-c)(z-d)}}=\frac{2}{\sqrt{(a-c)(b-d)}}F(\phi,k) \;,
\end{equation}
where $F(\phi,k)$ is the elliptic integral of the first kind,
\begin{equation}
F(\phi,k)=\int_0^\phi\frac{d\theta}{\sqrt{1-k\sin^2\theta}} \;.
\end{equation}
The second one is
\begin{eqnarray}
\lefteqn{\int\frac{dz}{(z-c)(z-d)}\sqrt{\frac{(z-a)(z-b)}{(z-c)(z-d)}}=\frac{2}{(c-d)^2}\left[\frac{(a-b)(c-d)}{\sqrt{(a-c)(b-d)}}F(\phi,k)\right. }\nonumber\\
&& \left. -2\sqrt{(a-c)(b-d)}E(\phi,k)+(z-a)\frac{(z-c)(b-d)+(z-d)(b-c)}{\sqrt{(z-a)(z-b)(z-c)(z-d)}}\right] \;,
\end{eqnarray}
where $E(\phi,k)$ is the elliptic integral of the second kind,
\begin{equation}
E(\phi,k)=\int_0^\phi d\theta\sqrt{1-k\sin^2\theta} \;.
\end{equation}
Finally, the integral for the distribution of the current density in
Eq.~(\ref{eq:form3}) is
\begin{eqnarray}
\lefteqn{\int\frac{dz}{z}\Big(\frac{z}{z-a}+\frac{z}{z-d}-\frac{b}{z-b}-\frac{c}{z-c}\Big)\sqrt{\dfrac{(z-a)(z-b)}{(z-c)(z-d)}}=\frac{4\sqrt{(a-c)(b-d)}}{c-d}E(\phi,k) } \nonumber\\
&& +\frac{4(a-b)}{\sqrt{(a-c)(b-d)}}\Big[\Pi(n,\phi,k)-\Pi(bn/a,\phi,k)\Big]-\frac{2(z-a)\big[(z-c)(b-d)+(z-d)(b-c)\big]}{(c-d)\sqrt{(z-a)(z-b)(z-c)(z-d)}} \;.
\end{eqnarray}

\twocolumn

\begin{thebibliography}{10}

\bibitem{Onsager49}
L. Onsager, Suppl. Nuovo Cimento {\bf 6},  249  (1949).

\bibitem{Feynman55}
R.~P. Feynman,  in {\em Progress in Low Temperature Physics}, edited by C.~J.
  Gorter (Elsevier Science Publishers B.V., Amsterdam, 1955), Vol.~1, Chap.~2,
  pp.\ 17--53.

\bibitem{Hatsuda94}
M. Hatsuda, S. Yahikozawa, P. Ao, and D.~J. Thouless, Phys. Rev. B {\bf 49},
  15870  (1994).

\bibitem{Lamb32}
S.~H. Lamb, {\em Hydrodynamics} (The University Press, Cambridge, 1932).

\bibitem{Donnelly91}
R.~J. Donnelly, {\em Quantized Vortices in Helium II} (Cambridge University
  Press, New York, 1991).

\bibitem{VonKlitzing80}
von Klitzing, K.~G. Dorda, and M. Pepper, Phys. Rev. Lett. {\bf 45},  494
  (1980).

\bibitem{Ashton79}
R.~A. Ashton and W.~L. Glaberson, Phys. Rev. Lett. {\bf 42},  1062  (1979).

\bibitem{Drew95}
H.~D. Drew and T.~C. Hsu, Phys. Rev. B {\bf 52},  9178  (1995).

\bibitem{Ichiguchi98}
T. Ichiguci, Phys. Rev. B {\bf 57},  638  (1998).

\bibitem{Hill53}
D.~L. Hill and J.~A. Wheeler, Phys. Rev. {\bf 89},  1102  (1953).

\bibitem{Peierls57}
R.~E. Peierls and J. Yoccoz, Proc. Roy. Soc. {\bf A 70},  381  (1957).

\bibitem{Peierls62}
R.~E. Peierls and D.~J. Thouless, Nucl. Phys. {\bf 38},  154  (1962).

\bibitem{Ortiz95}
G. Ortiz and D.~M. Ceperley, Phys. Rev. Lett. {\bf 75},  4642  (1995).

\bibitem{Giorgini96}
S. Giorgini, J. Boronat, and J. Casulleras, Phys. Rev. Lett. {\bf 77},  2754
  (1996).

\bibitem{Vitiello96}
S.~A. Vitiello, L. Reatto, G.~V. Chester, and M.~H. Kalos, Phys. Rev. B {\bf
  54},  1205  (1996).

\bibitem{Sadd97}
M. Sadd, G.~V. Chester, and L. Reatto, Phys. Rev. Lett. {\bf 79},  2490
  (1997).

\bibitem{Feynman53}
R.~P. Feynman, Phys. Rev. {\bf 91},  1301  (1953).

\bibitem{Chester68}
G.~V. Chester, R. Metz, and L. Reatto, Phys. Rev. {\bf 175},  275  (1968).

\bibitem{Feynman56}
R.~P. Feynman and M. Cohen, Phys. Rev. {\bf 102},  1189  (1956).

\bibitem{Saffman95}
P.~G. Saffman, {\em Vortex Dynamics} (Cambridge University Press, New York,
  1995).

\bibitem{Feynman54}
R.~P. Feynman, Phys. Rev. {\bf 94},  262  (1954).

\bibitem{Clements93}
B.~E. Clements, J.~L. Epstein, E. Krotscheck, and M. Saarela, Phys. Rev. B {\bf
  48},  7450  (1993).

\bibitem{Fetter71}
A.~L. Fetter, Phys. Rev. Lett. {\bf 27},  986  (1971).

\bibitem{Duan94}
J.~M. Duan, Phys. Rev. B {\bf 49},  12381  (1994).

\bibitem{Duan95}
J.~M. Duan, Phys. Rev. Lett. {\bf 75},  974  (1995).

\bibitem{Niu94}
Q. Niu, P. Ao, and D.~J. Thouless, Phys. Rev. Lett. {\bf 72},  1706  (1994).

\bibitem{Niu95}
Q. Niu, P. Ao, and D.~J. Thouless, Phys. Rev. Lett. {\bf 75},  975  (1995).

\bibitem{Landau77}
L.~D. Landau and E.~M. Lifshitz, {\em Quantum Mechanics: Non-Relativistic
  Theory}, 3rd ed. (Pergamon, Oxford, 1977).

\end{thebibliography}
\end{document}